# Guidelines for the stabilization of a polar rhombohedral phase in epitaxial Hf$_{0.5}$Zr$_{0.5}$O$_2$ thin films

Pavan Nukala, Yingfen Wei, Vincent de Haas, Qikai Guo, Jordi Antoja-Lleonart & Beatriz Noheda

Published online: 22 Dec 2020.

Submit your article to this journal ↗

Article views: 47

View related articles ↗

View Crossmark data ↗

# Guidelines for the stabilization of a polar rhombohedral phase in epitaxial $Hf_{0.5}Zr_{0.5}O_2$ thin films

Pavan Nukala[†], Yingfen Wei[†], Vincent de Haas, Qikai Guo, Jordi Antoja-Lleonart, and Beatriz Noheda

Zernike Institute of Advanced Materials, University of Groningen, Groningen, The Netherlands

**ABSTRACT**

The unconventional Si-compatible ferroelectricity in hafnia-based systems, which becomes robust only at nanoscopic sizes, has attracted a lot of interest. While a metastable polar orthorhombic (o-) phase ($Pca2_1$) is widely regarded as the responsible phase for ferroelectricity, a higher energy polar rhombohedral (r-) phase is recently reported on epitaxial $HfZrO_4$ (HZO) films grown on (001) $SrTiO_3$ ($R3m$ or $R3$), (0001) GaN ($R3$), and Si (111). Armed with results on these systems, here we report a systematic study leading toward identifying comprehensive global trends for stabilizing r-phase polymorphs in epitaxially grown HZO thin films (6 nm) on various substrates (perovskites, hexagonal and Si).

## 1. Introduction

Ferroelectric hafnia-based compounds owing to their Si-compatibility are very promising materials that can seamlessly integrate ferroelectric phenomena into microelectronic devices [1–3]. In these systems, ferroelectricity becomes more robust with device miniaturization, quite contrary to the behavior of conventional ferroelectrics where depolarization fields become increasingly important at small sizes [4–6]. Hence this is a new kind of ferroelectricity, leading to a growing interest in not just application-oriented research but also in fundamental research on its origins and features [7–17].

Hafnia (zirconia) and hafnia-based alloys characteristically display a plethora of polymorphs [18]. While the monoclinic ($P2_1/c$, m-) phase is the bulk ground state, other low-volume metastable polymorphs such as the tetragonal (t-), cubic (c-) or orthorhombic (o-) phases are responsible for the various functionalities in these materials [19,20]. These are high temperature, high pressure phases in the bulk, which can be stabilized at ambient conditions via nanostructuring [21], doping [1,10,22–25], oxygen-vacancy engineering [26,27], thermal stresses [28,29] and epitaxial strain [22,25,30–37], all of which can be suitably factored into thin film geometries. In particular, ferroelectric behavior results from the metastable polar phases. First-principles structure calculations predict that at least five polar polymorphs (with space groups $Pca2_1$, $Cc$, $Pmn2_1$, $R3$ and *R3m*) fall in an energy window possible to achieve experimentally, among which the polar o-phase (*Pca2₁*) has the least energy (64 meV/f.u.) [30,38,39]. This phase has been observed (and sometimes assumed) in polycrystalline ferroelectric layers grown via atomic layer deposition (ALD) [1,10,40,41], chemical solution deposition (CSD) [42], RF sputtering [43,44], co-evaporation and plasma assisted atomic oxygen deposition [40], as well as epitaxial layers obtained via pulsed laser deposition (PLD) [25,33–35,37,45,46]. Recently, a higher energy polar rhombohedral (r-) phase has been observed on $Hf_{0.5}Zr_{0.5}O_2$ (HZO) layers epitaxially grown on $SrTiO_3$ (STO) substrates buffered with $La_{0.7}Sr_{0.3}MnO_3$ (LSMO) as the back-electrode [30]. Films below 9 nm exhibit single r-phase, which is characterized by wake up-free polarization switching. Remanent polarization ($P_r$) values as large as 34 μC/cm² were observed on these films with thickness of 5 nm, this being the highest $P_r$ reported in $HfO_2$-$ZrO_2$ alloys. Later, pure r- phase was also reported on 6 nm HZO layers grown epitaxially on hexagonal GaN buffered Si substrates [47], whereas mixed m- and r-phases are reported on HZO layers epitaxially grown directly on Si (111) [48].

The experimental demonstration of these reported polar and in most cases ferroelectric o- and r- phases, begs the question of which factors favor which phase. By systematically varying the initial strain conditions and film orientation through a choice of various substrates (using PLD), we present here a comprehensive study leading to guidelines on the stabilization of the r-phase polymorphs. In deriving trends we utilize the results recently reported for epitaxially grown 6 nm HZO layers on STO//LSMO (001) [9], hexagonal GaN (0001) [47], and Si (111)) [48], in addition to new data acquired in this work on other (001) perovskite substrates and (0001) hexagonal sapphire.

## 2. Experimental methods

HZO thin films of thickness 6 nm were grown by PLD on LSMO-buffered perovskite (001) substrates (in the following denoted as LBP), Nb (0.5%) doped $SrTiO_3$ (Nb:STO, (001)) and hexagonal sapphire (0001). A KrF excimer laser ($\lambda = 248$ nm) was used for ablation. The HZO target was synthesized via standard solid-state synthesis (sintering temperature: 1400 °C), starting from powders of $HfO_2$ (99%) and $ZrO_2$ (99.5% purity). LSMO targets were purchased from PI-KEM. For HZO layers on LBP, an LSMO layer was first deposited as the bottom electrode on various perovskite substrates ($YAlO_3$, $LaAlO_3$, $NdGaO_3$, $(LaAlO_3)_{0.3}(Sr_2TaAlO6)_{0.7}$, $SrTiO_3$ and $DyScO_3$) at a laser fluence and frequency of 1 J cm⁻², and 1 Hz, respectively, with a 0.15 mbar of oxygen pressure and substrate temperature of 775 °C, giving rise to a growth rate of ~0.13 Å/s. This LSMO buffer layer had a thickness of $t = 40$ nm, unless otherwise mentioned. The HZO layers on LBP, Nb doped STO and sapphire were deposited at 1.1 J cm⁻², 2 Hz, 0.1 mbar and 800 °C (deposition rate: ~0.16 Å/s). Films were cooled down at 5 °C/min to room temperature under oxygen pressure of 300 millibar.

Global structure, symmetry, phase-mixing and domains information was obtained from X-ray diffraction (Cu Kα source). Texture analysis was performed via χ-φ (pole-figure) scans at $2\theta \approx 30.0°$ (approximately corresponding to the $d_{\{111\}}$ of the low-volume phases, including c-, t-, o-, r-phases), and at $2\theta \approx 34.5°$ (approximately corresponding to the $d_{\{200\}}$ of all the polymorphs). These will be referred to as {111} pole figure and

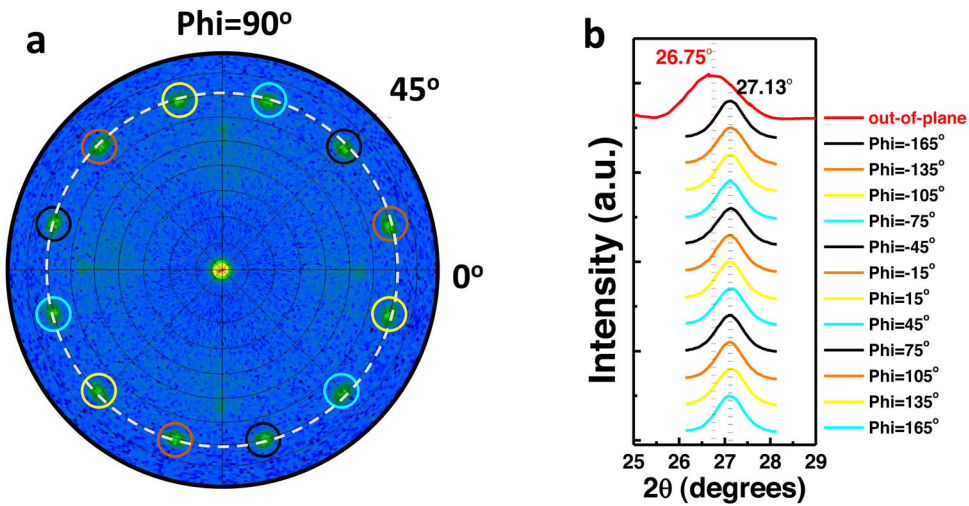

**Figure 1.** R-phase determination in HZO//LBP, P = STO. (a) Pole-figures or $\chi$-$\varphi$ projections obtained at a fixed $2\theta$ corresponding to $d_{(111)}$ of HZO films (9 nm). (b) $2\theta$ scans about all the 13 poles in (a). Reproduced with permission from Ref. [30].

{001} pole figures, respectively. The d-spacings of the poles obtained from the $\chi$-$\varphi$ projections were more precisely analyzed through symmetric $2\theta$ scans around them.

Local structural characterization and phase analysis was performed through STEM imaging at 300 kV (Titan G2, and Themis). STEM images were obtained in high-angle annular dark field (HAADF) mode. EEL spectra were obtained from Gatan Enfinium spectrometer with simultaneous dark-field STEM atomic resolution imaging. O–K edge spectra were obtained at an energy dispersion of 0,25 eV, exposure time of 0.5 s integrated over 50 acquisitions (parallel to the interface).

# 3. Results and discussion

## 3.1. Discovery of r-phase on HZO

Figure 1(a,b) shows the {111} pole figure of a HZO film epitaxially grown on LSMO-buffered STO and corresponding $2\theta$ scans about each pole, reported by Wei et al. [17] This 111 reflection was observed at about $2\theta \approx 27°$ with synchrotron radiation of $\lambda = 1.378$ Å, corresponding to $2\theta \approx 30°$ in the lab source with $\lambda = 1.541$ Å. The pole figure clearly shows that the HZO films are (111) oriented (Figure 1(a)). Presence of 12 poles at $\chi = 71°$ is a result of four domains rotated from each other by 90° with respect to the film normal, with each domain contributing to three poles i.e., (11−1), (−111) and (1−11), separated by $\varphi = 120°$. Figure 1(b) clearly shows that the 12 poles at $\chi = 71°$ share the same $2\theta$ value, which is larger than that of the pole at $\chi = 0°$ (out-of-plane). This indicates a 3:1 multiplicity in the d-spacing of the {111} planes, with the longer lattice parameter out of plane ($d_{(111)}$=2.98 Å > $d_{(11−1)/(1−11)/(−111)}$=2.94 Å), a signature of rhombohedral symmetry. In contrast, the other low volume polymorphs, such as c-, t-, o-phase, exhibit the same d-spacing for all the {111} planes.

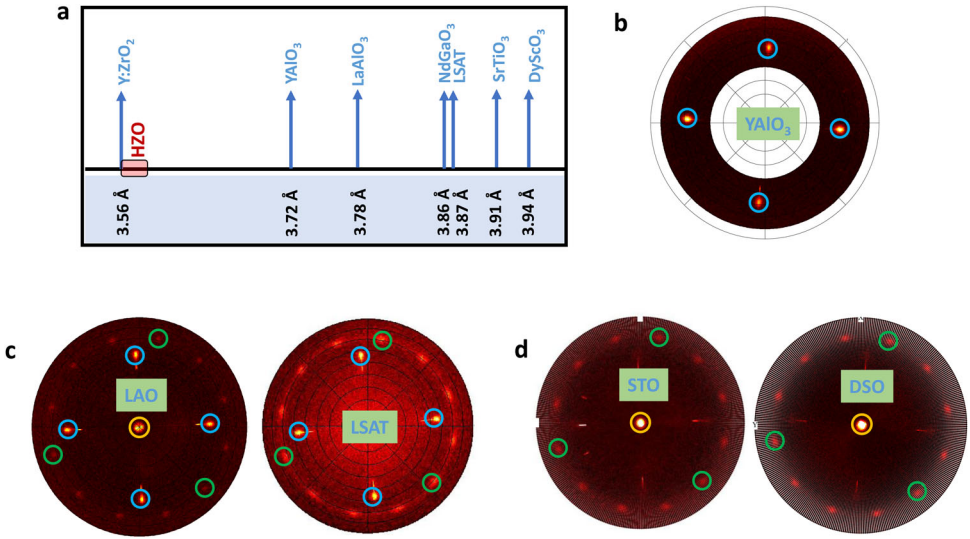

**Figure 2.** Texture measurements of HZO films grown on various LSMO buffered (001) perovskite substrates. (a) List of perovskite substrates used in this work ordered by their (pseudo)cubic lattice parameter. (b–d) Pole figures obtained at $2\theta \approx 30°$, showing an (001) growth orientation on YAO (b), mixed (001) and (111) orientation on LAO and LSAT (c), and a unique (111) orientation on STO and DSO substrates.

### 3.2. HZO on LSMO-buffered perovskites (001): polar r-phase vs. non-polar phases

#### 3.2.1. Orientation changes with increasing substrates lattice parameters

The list of LBP substrates used, and their corresponding pseudo-cubic lattice parameters are shown in Figure 2(a). The relevant lattice parameters $d_{\{110\}}$ and $d_{\{1-10\}}$ of HZO lie between 3.56 Å and 3.62 Å for various polymorphs. Thus, all the LBP substrates provide an initial tensile strain for a cube-on-cube growth.

{111} pole-figures of HZO//LBP with P = YAlO₃ (YAO, $a = 3.72$ Å), show 4 intense poles (blue circles) appearing at $\chi \approx 57°$, separated in $\varphi$ by 90° (Figure 2(b)). This can solely arise from a film completely oriented along (001). Pole figures of HZO on LBP with P = LaAlO₃ (LAO, $a = 3.78$ Å) and (LaAlO₃)₀.₃(Sr₂AlTaO₆)₀.₇ (LSAT, $a = 3.87$ Å) are shown in Figure 2(c). In addition to the four intense poles at $\chi \approx 57°$, 12 weaker poles appear at $\chi \approx 71°$ (green circles). These additional poles arise from (111) oriented grains with four in-plane domains, analogous to the case of HZO//LBP with P = STO [17]. Thus, HZO on these substrates exhibits a mixed (001) and (111) oriented grains. The domains arising from these two different orientations will be referred to as D-(001) and D$i$-(111) with $i = 1–4$, respectively. HZO layers on LBPs with P = STO ($a = 3.91$ Å) and DyScO₃ (DSO, $a_{\text{psudeocubic}} = 3.94$ Å) exhibit single (111) orientation (Figure 2(d)). Thus, increasing the perovskite lattice parameter results in a textural transformation of the HZO layers from purely (100)-oriented to purely (111)-oriented, proceeding through a mixed orientation.

#### 3.2.2. Phase analysis

We further analyzed the domains in the different film textures to determine the phases comprising them. Here we present this analysis on a representative example of HZO//

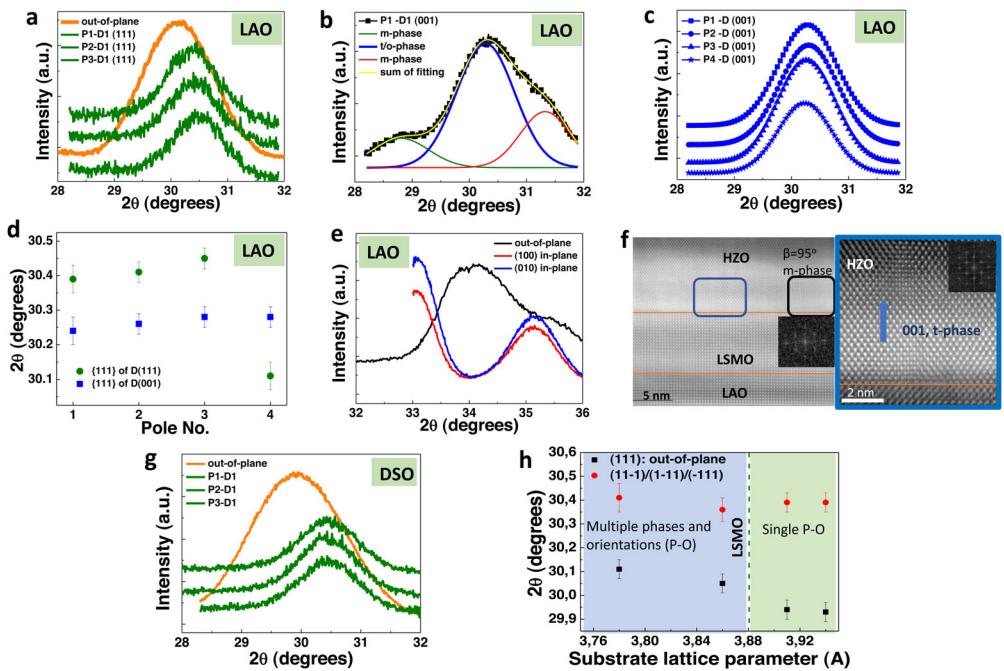

**Figure 3.** Phase determination of HZO//LBP, P = LAO and DSO. (a) $2\theta$ scans about the all the four {111} poles of a representative domain D1 in a (111)-oriented grain of HZO//LBP, P = LAO. (b) Three peak Gaussian fitting of the $2\theta$ scans about each of the {111} poles of (001) oriented grains in HZO// LBP, P = LAO. (c) $2\theta$ scans about the all the four {111} poles of (001)-oriented grain of HZO//LBP, P = LAO. (d) $2\theta$ peak positions of the {111} poles in a representative domain of (111) oriented grain (green), compared with those of the low-volume phase of the (001) oriented grain (blue). (e) $2\theta$ scans about each of the {200} poles of the (001) oriented grains in HZO//LBP, P = LAO. (f) HAADF-STEM image showing a mixture of (001)-oriented m- and t-phase regions (and their Fourier transforms) on HZO//LBP, P = LAO. (g) $2\theta$ scans about the all the four {111} poles of a representative domain of HZO//LBP, P = DSO. (h) Comparison between $d_{(111)}$ and $d_{(11-1)}$ of the (111)-oriented rhombohedral grains on various substrates. LSMO lattice parameter is indicated by the vertical dashed line. Single phase and orientation are obtained when LSMO is tensile strained.

LBP, with P = LAO. Note that these films contain both D-(001) and Di-(111). Three poles at $\chi \approx 71°$ of representative domain D1-(111) (P1–P3 indicated in green in Figure 2(c)) have a peak at a larger $2\theta$ (=30.4°) than the pole at $\chi \approx 0°$ ($2\theta = 30.1°$)) (Figure 3a, d-green). Such a clear 3:1 multiplicity in the d-spacing of the {111} planes is only consistent with the rhombohedral symmetry, as discussed above, for the (111)-oriented grains (see Figure 1(b)). Representative $2\theta$ scan from the poles of D-(001) (P1–P4, indicated in blue circles in Figure 2(b)) is shown in Figure 3(b). Scans from each pole can be deconvoluted into three Gaussians, with the $2\theta$ positions of the extreme peaks consistent with the $d_{{111}}$ of the bulk m-phase (at $2\theta = 28.6°$, green and $2\theta = 31.4°$, red in Figure 3(b)). The middle peak (shown in blue line) corresponds to the $d_{{111}}$ of a low-volume phase. As shown in Figure 3(c) (and Figure 3(d), blue), this low-volume phase peak has the same $2\theta$ position for all the poles (within the error of estimation), which is characteristic of one of t- or c- or o-phases. Figure 3(d) shows a comparison of $d_{{111}}$ of the rhombohedral (111) (green) and the (001) oriented grains (blue).

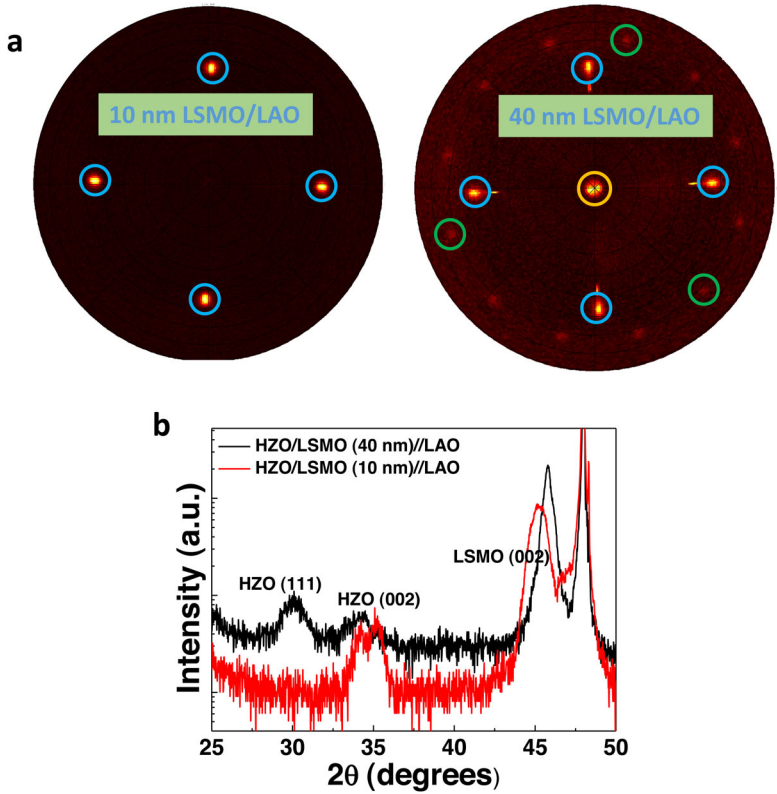

**Figure 4.** Texture and phases of HZO//LBP, P = LAO by varying the LSMO thickness. (a) Comparison of {111} pole figures of HZO//LBP, P = LAO at LSMO thickness of 10 nm and 40 nm. (b) Specular $\theta$–$2\theta$ reflections comparisons between the two, revealing that while 10 nm LSMO layer is fully strained, the 40 nm one is strain relaxed.

Further identification of the precise phase was performed from the $2\theta$ scans about the {001} poles. Figure 3(e) shows that the d-spacings of the {002} planes of the low volume phase (deconvoluted from the monoclinic phase) are identical ($d_{002} = d_{200} = d_{020} = 2.55$ Å). These parameters are more in line with the non-polar t-phase, than with the o-phase (with $d_{\{200\}} = 2.51$, 2.55 and 2.62 Å). Thus, (001)-oriented grains exhibit a mixture of non-polar m- and t-phases. A minor amount of (polar) o-phase, as suggested by Yoong et al. [35] cannot however be discounted. HAADF-STEM analysis on HZO//LBP, P = LAO further confirms the co-existence of t-phase and m-phase in (001)-oriented domains. In Figure 3(f) (left panel), the region corresponding to m-phase ($\beta = 95°$) and the region corresponding to the low-volume phase, are marked in black and blue, respectively. The low-volume region (in blue) has predominantly t-phase (as deduced from the FFT in the inset of Figure 3(d) right panel). Thus, while the (001)-oriented domains exhibit non-polar phases (m-, t-phase), the (111)-oriented domains correspond to a polar r-phase.

The correlation between the film orientation and the corresponding phases observed on LAO is consistent across other substrates too, e.g. $2\theta$ scans performed about {111} and {001} poles, on (001)-only oriented HZO on LBP, P = YAO exposes a mixture of non-polar t and m-phases (data not shown), while $2\theta$ scans from (111)-only oriented

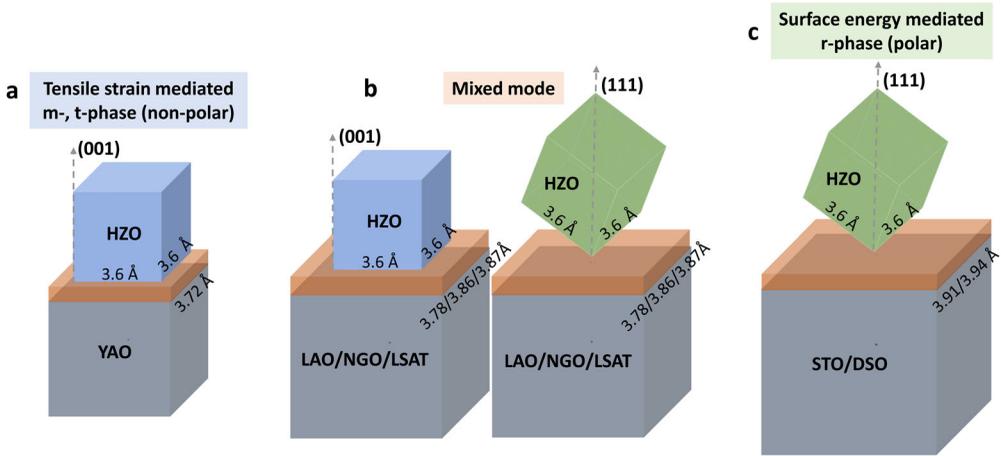

**Figure 5.** Model for growth of HZO layers on LBP. (a) A cube-on-cube growth mode ((001)-orientation) determined by initial tensile strain conditions of the substrate is observed at modest mismatch between the lattice parameters of substrate (e.g. YAO) and the film (see Figure 2(a)). As substrate mismatch increases, growth proceeds toward a surface energy mediated mode. (b) A mixed mode of growth on LBP, P = LAO, NGO, LSAT resulting in both (001) and (111) oriented grains. (001) grains exhibit non-polar m- and t-phases, (111) exhibit polar r-phase. (c) On LBP, P = STO and DSO unique orientation and a single r-phase is obtained, and the growth is purely surface-energy mediated.

HZO on LBP, P = STO (Figure 1(a)) or DSO (Figure 3(g)) expose a pure r-phase with 3:1 multiplicity of the {111} planes.

Interestingly, the strain state of the back-electrode, LSMO, also influences the phase and orientation of HZO layer. Tensile strained LSMO layers (substrate lattice parameter >3.88 Å) promote a pure r-phase oriented along (111) (green region in Figure 3(h)). Estandia et al. [49], studied HZO (9 nm)/LBP layers with P = TbScO$_3$, GdScO$_3$ and NdScO$_3$. These substrates have larger lattice parameters than both STO and DSO (thus larger tensile strain on LSMO) and also expose (111) oriented HZO layers, with single polar phase, quite consistent with our analysis [49]. However, when LSMO is compressively strained (blue region in Figure 3(h)), (001) oriented non-polar m and t-phases, always appear. At lower values of compressive strains on LSMO (~3%, substrates: LSAT, NGO, LAO), the (111)-oriented polar r-phase coexists with these non-polar phases, but at larger values (> 4%, substrate: YAO)), the (111) oriented r-phase completely disappears, and HZO layers stabilize solely in (001) oriented non-polar phases.

We further substantiated the effect of compressively strained LSMO layers on the phase and orientation of HZO by varying the thickness of LSMO. Our 6 nm thick HZO layer grown on a 40 nm thick (partially relaxed) LSMO layer, yields a combination of (001) and (111) oriented domains (Figure 4(a) right panel); while a thinner, fully strained, LSMO layer (t = 10 nm) yields only (001) oriented HZO films (Figure 4(a) left panel). Figure 4(b) clearly shows that the out of plane d$_{(002) pc}$= 1.98 Å (1.8% larger than bulk) on the 40 nm thick LSMO, whereas d$_{(002) pc}$=2.00 Å (2.8% larger than bulk) on the10 nm thick layer. This is a result of a larger in-plane compressive strain in the 10 nm LSMO layer, consistent with the trends shown in Figure 3(h).

### 3.2.3. Growth model for HZO layers

Our results on various LBPs allows us to propose the following model for the growth of HZO layers (Figure 5). With P = YAO, HZO layers grow in a cube-on-cube fashion resulting in (001) oriented layers (Figure 5(a)). In this setting YAO offers initial tensile strain boundary conditions of ~3.3% to the (001) oriented HZO layers. Upon progressing to substrates with larger lattice parameters (LAO, NGO, LSAT), the tensile strain for a cube on cube growth mode increases (>5%), and domains with (111) orientation also start appearing (Figure 5(b)). This is because strain can no longer mediate the cube-on-cube growth, and other mechanisms come into play. In $HfO_2$ and $ZrO_2$ particles, it is well known that the (111) surfaces have the least energy [21]. Thus, the transition from (001) oriented films to (111) oriented films corresponds to a transition from tensile-strain mediated growth (Figure 5(a)) to a surface energy mediated growth (Figure 5(c)). The (001) oriented domains predominantly crystallize in non-polar monoclinic and tetragonal phases. However, the (111) oriented domains stabilize in a polar r-phase. A pure r-phase with single (111) orientation can be preferentially stabilized upon further increasing the substrate lattice parameter (STO, DSO). As proposed by Wei et al. [30], nanoparticle pressure (surface energy induced pressure from small domain sizes) stabilizes the (111)-oriented cubic phase at high temperatures (growth temperature). The thermal expansion coefficient mismatch between the substrate and HZO layers[1] provides an added biaxial compressive strain to HZO (111) layers, stabilizing the r-phase at room temperature [50,51]. Furthermore, as recently proposed by Estandia et al. [52], domain matching epitaxy mechanism may further assist the surface energy-mediated growth of HZO layers for efficient strain relaxation.

The stabilization of a pure polar phase requires effective screening of the depolarization fields, and this brings in the role of the back-electrode (LSMO). On substrates such as STO and DSO, LSMO is tensile strained exposing a single polar phase for HZO. However, when LSMO is compressively strained, non-polar phases always appear. To further glean into the role of the strain state of LSMO, we compared the interfaces and depolarization mechanisms of HZO layers on LBP with P = STO and LAO.

Wei et al. [30], have reported the existence of an interfacial tetragonal phase between LSMO and HZO in HZO//LBP (P = STO) (Figure 6(a)), something which is absent in the case of HZO//LBP (P = LAO) (Figure 6(b)). Electron energy loss spectroscopy (EELS) analysis of the O-K edge (Figure 6(c) upon normalizing with the thickness of the sample) clearly shows that this interface is oxygen deficient compared to the rest of r-HZO. It is well-known that tensile strain conditions promote the formation of oxygen vacancies ($V_o$) in perovskites films [53]. The correlation between the oxygen deficient HZO interface and tensile strained LSMO, strongly suggests that it is the $V_o$ in the latter that are responsible for the formation of such an interface. Such a mechanism of $V_o$ transfer between various layers is well-reported in several interfacial memristive systems involving manganites and nickelates [54–56]. From the first-principles calculations of Rushchanskii and coworkers, [57] these oxygen-deficient tetragonal phases in $HfO_2$ and $ZrO_2$ can be conducting, yielding an additional screening mechanism in for the stabilization of pure r-phase [58,59].

---

[1]Thermal expansion coefficient (α) of STO is $11.1 \times 10^{-6}$/°C from room temperature to 1000 °C, and that of cubic phase Yttria Stabilized Zirconia (YSZ, sister compound of HZO) is $8.5 \times 10^{-6}$/°C from 60 to 900 °C. $\alpha_{YSZ}$ is also found to be independent of Yttria concentration [50,51].

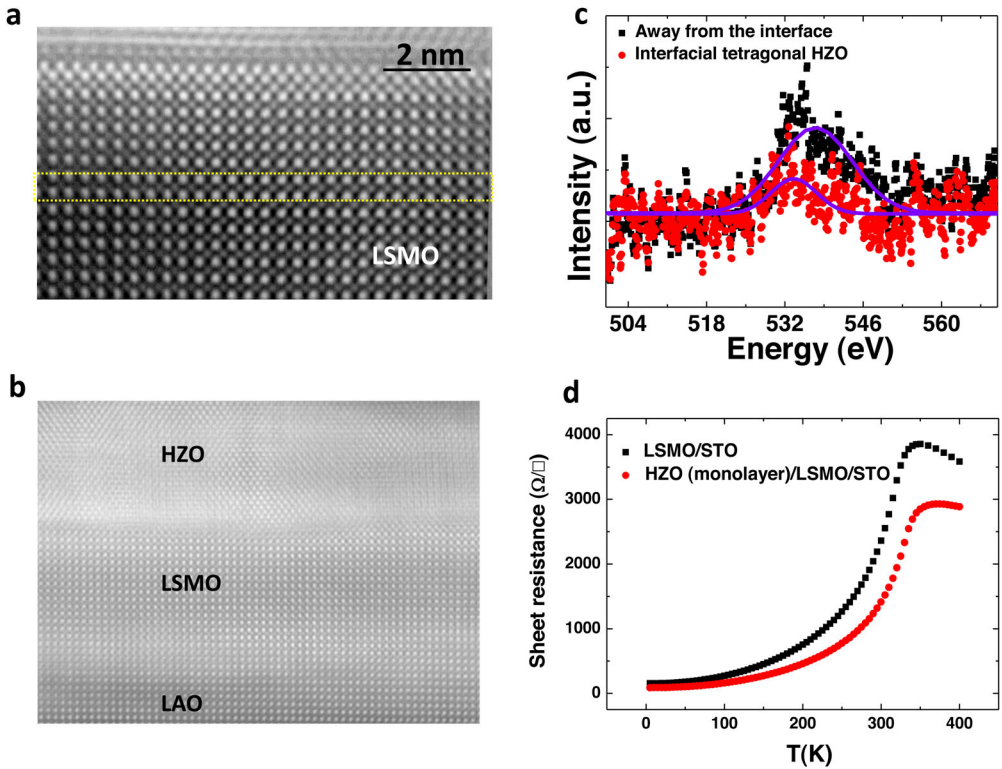

**Figure 6.** Interfacial layer characterization between tensile strained LSMO and HZO. (a) HAADF-STEM image of the interfacial layer between LSMO and HZO (substrate is STO). Reproduced with permission from Ref [30]. (b) HAADF-STEM image of HZO//LBP, P = LAO. LSMO (10 nm) is compressively strained and exhibits no interfacial layer. (c) EELS O k-edge (after background subtraction, deconvolution of plural scattering, and normalization to the continuum), comparison between the interfacial layer and the HZO- away from the interface in HZO//LBP, P = STO. (d) Sheet resistance-temperature characteristics from 5 to 400 K of bare LSMO//STO with HZO (monolayer)/LSMO//STO. HZO monolayer scavenges oxygen vacancies from LSMO, rendering LSMO more conducting in the measured temperature range.

Furthermore, transfer of $V_o$ to the interface reduces the defect content of LSMO, enhancing the overall conductivity and the electrostatic screening. The larger sheet resistance, measured using the van der Pauw method, of just the LSMO//STO layers compared to the HZO(monolayer)/LSMO//STO samples (Figure 6(d)) is consistent with this hypothesis. In contrast, in compressively (and fully) strained LSMO (e.g. on LAO), $V_o$ formation is hindered, resulting in no conducting interfacial phase, limiting the screening mechanisms that can stabilize the polar phase (Figure 6(b)).

### 3.3. HZO on Nb (0.5%) doped STO

The {111} pole figure shown in Figure 7(a) reveals that HZO layers that are grown on Nb:STO have their [111] direction mis-oriented by about 15° with respect to the surface normal. The substrate imposes its four-fold symmetry on the film resulting in four ferroelastic domains rotated about the surface normal by 90° with respect to each other. It can be shown quite rigorously that such tetra-domain structure accurately reproduces all the χ- φ poles observed in Figure 7(a) (poles from every domain are encircled in a different color). 2θ scans across the poles in a representative domain are shown in Figure 7(b). It can be seen that there are three different {111} d-spacings, which rules out a rhombohedral phase and it is, instead, characteristic of the triclinic symmetry. These results clearly evidence that the r-phase formation is strongly correlated to the film orientation being (111), with any deviations resulting in the lowering of symmetry. These results also point out the importance of the oxygen-deficient interfacial layer in LSMO-buffered STO substrates in stabilizing the (111) orientation and, subsequently, the r-phase on perovskites.

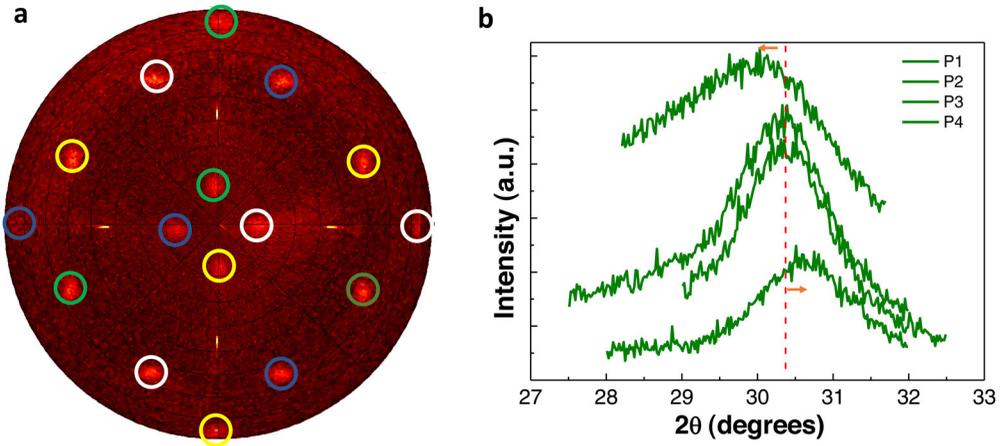

**Figure 7.** Texture and phase of HZO layers on Nb (0.5%)-doped STO. (a) {111} pole figures of HZO layers on Nb:STO. Four domains with (111) in each domain misoriented from the surface normal by 15° can be observed. (b) 2θ scans about the all the four {111} poles of a representative domain. The multiplicity of the $d_{\{111\}}$ is consistent with a lower symmetry triclinic phase.

### 3.4. HZO on hexagonal and trigonal substrates: polar r-phase

Next, we grew 6 nm thick HZO layers on substrates that, by virtue of their symmetry, impose (111) orientation to the film. In this pursuit, we used (0001) hexagonal substrates such as GaN buffered Si [47] with $a_1 = a_2 = 3.23$ Å, α = 120° and sapphire with $a_1 = a_2 = 3.46$ Å, α = 120°, both of which provide an initial compressive strain to the {111} plane of HZO ($a_1$, $a_2 \sim 3.56$–3.62 Å, α $\sim$ 120° depending on the polymorph). The {111} pole figures on both these substrates (Figure 8(a,b)) clearly show that these films are (111) oriented. Furthermore, there are six {11−1} poles at χ = 71°, separated in φ by 60°, arising out of two domains ($D_h1$ and $D_h2$) 180° rotated with respect to the film normal. Phase analysis by means of 2θ scans around poles corresponding to a representative domain (Figure 8(c,d)) reveals 3:1 multiplicity, pertaining to an r-phase, on both these substrates. HAADF-STEM images reported by Lours et al. [47] clearly show these domains and the coherent domain boundaries on the GaN-buffered Si substrate (Figure 8(e)). The HAADF-STEM image from just one domain (Figure 8(f)) shows cationic columns of alternating intensity and shape along the [112] (in-plane) direction. This is characteristic of the r-phase, and it is not found in any other low-volume phases, as illustrated in the inset of Figure 8(f) with a multislice simulation of a 20 nm thick cross-sectional lamella. Thus, both XRD and electron microscopy independently confirm the *r*-phase symmetry.

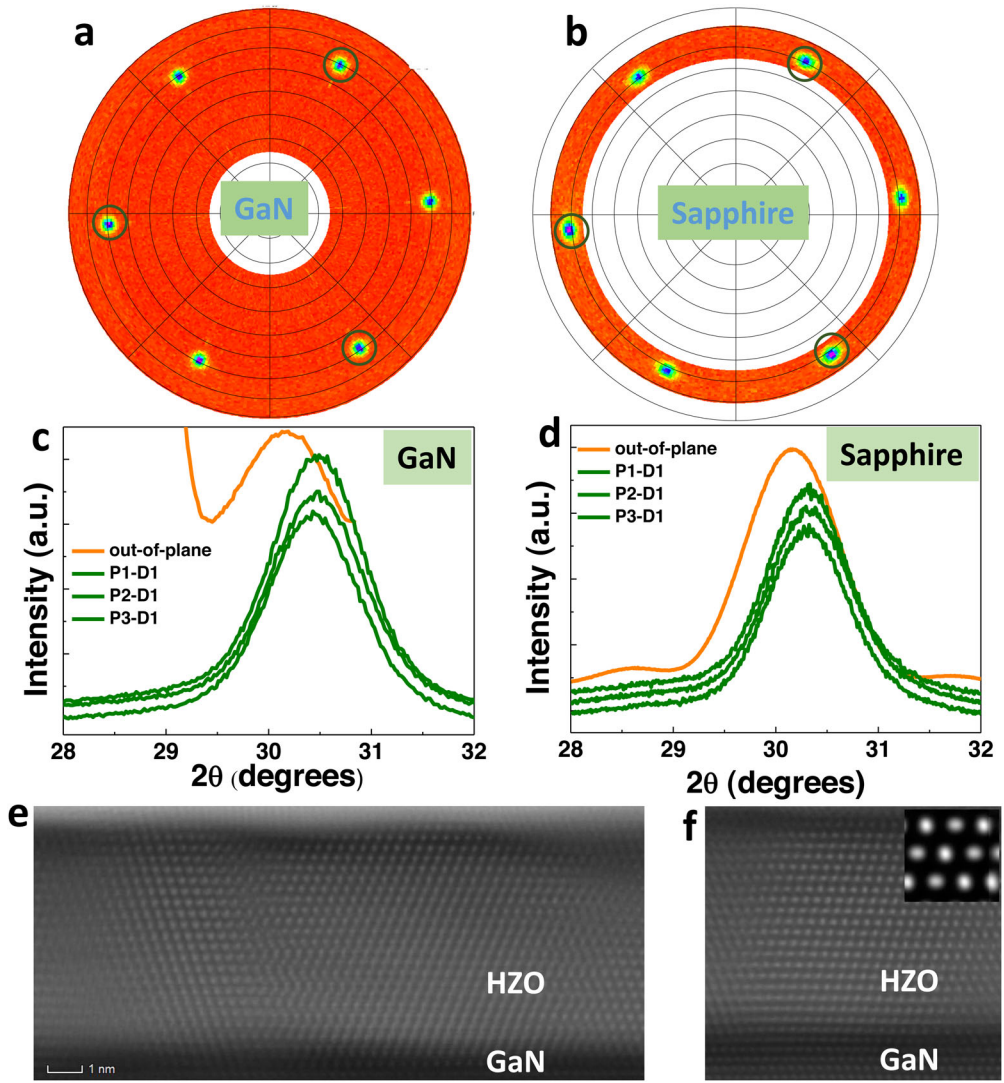

**Figure 8.** Characterization of HZO layers on (0001) hexagonal substrates. Pole-figures of HZO layers on GaN, and sapphire (b). $2\theta$ scans about the all the four {111} poles of a representative domain of HZO on GaN (buffered Si) (c), and sapphire (d). HAADF-STEM image of 180° domain boundary of HZO//GaN buffered Si (e), and of a single domain (f) consistent with r-phase symmetry (simulation in the inset). Part labels a, c, e and f are reproduced with permission from Ref. [47].

To further understand the precise symmetry of the r-phase on GaN, Lours et al. [60] performed differential phase contrast (DPC) STEM imaging on HZO layers on GaN-buffered Si substrate. The differential DPC (dDPC) images were simulated through multislice simulations on both *R3m* and *R3* (Figure 9(a)) structures at different lamella thicknesses. An experimental match in terms of the oxygen columnar positions was found with the *R3* phase (Figure 9(b)). Visually, this can be understood by looking at the O–Hf–O//O–Hf–O angle in both the structures (as indicated by the red and green lines in Figure 8(a,b)). While in the *R3m* structure they are collinear, in the *R3* structure they are not similar to the experimental image. Finally, from these dDPC images, by estimating the center of mass of cationic columns and anionic columns, Lours et al., reported $P_r$ values of 1.6–1.9 $\mu C/cm^2$, corresponding to a displacement vector of 8.5–9.0 pm/unit cell along [111] (Figure 9(c,d)). This is an order of magnitude less than the values measured on HZO//LBP, P = STO.

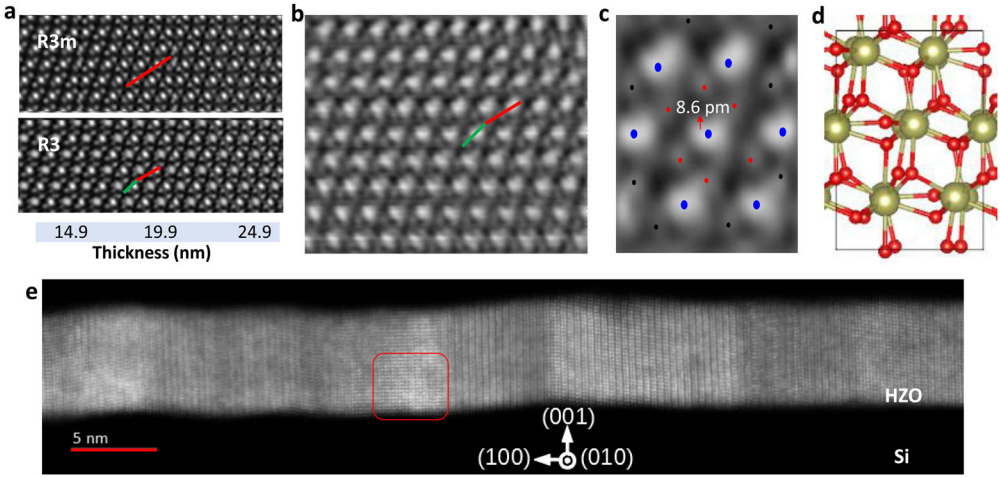

**Figure 9.** High resolution TEM analysis of polar r-phase and o-phase. (a) Multislice dDPC (differentiated differential phase-contrast) STEM image simulations of *R3m* and *R3* phases of HZO. (b) iDPC-STEM experimental image of the HZO layers on GaN buffered Si, corresponding to *R3* symmetry. (c) Displacement between centers of masses of positively (Hf/Zr) charged and negatively charged (O) columns of a representative unit cell. The [111] direction of this dipole moment is a further evidence of rhombohedral distortion. (d) Schematic HZO unit cell. (a–d) are reproduced with permission from Ref. [47] (e) HAADF-STEM image of the domain structure of HZO (100) grown with direct epitaxy on Si(100). The accordion shape is a result of the monoclinic domains, and some regions between these monoclinic domains show orthorhombic distortion. Reproduced from Ref. [48].

Nukala et al. [48], have synthesized epitaxial layers of ferroelectric HZO directly on Si. First on Si (111) substrates, they obtain a phase mixture of m-phase and polar r-phase. Very interestingly, regions of HZO which are directly in contact with the monolayer of β-cristobalite (c-SiO₂ phase) on Si (111) stabilize in an *r*-phase, while an m-phase results if amorphous SiO_x layer regrows at the interface. The (111) surface of β-cristobalite provides a hexagonal template and a small initial compressive strain conditions ($a_1=a_2=3.55$ Å, $\alpha = 120°$) for the growth of (111) HZO. Thus, HZO grown on hexagonal GaN-buffered Si, sapphire and trigonal Si (111), further reinforces the trends observed on LBP substrates, i.e. that a combination of initial compressive strain and the (111) orientation stabilizes the *r*-phase.

## 3.5. Polar o-phase

Since the *r*-phase with polarization direction along [111] can be stabilized in (111) oriented films, the question remains whether the polar o-phase with polarization along the c-axis can be stabilized for (001) oriented films. 001) HZO layers on LBPs shown in this work were predominantly stabilized as non-polar phases, owing to a combination of tensile strain from the substrate and bad screening from the LSMO. Nukala et al. [48] have shown the existence of the polar o-phase in combination with bulk m-phase in HZO layers epitaxially grown on Si (100) (Figure 9(e), boxed in red). Crystalline SiO$_2$ layer on the surface of Si offers an initial slight compressive strain condition for the growth of the HZO layers. However, due to the presence of a regrown amorphous SiO$_x$ at the interface, the authors suggest that it is unlikely that any substrate strain is transferred to the film. It appears that the stabilization of the polar o-phase (red box in Figure 9(c)) is a result of the inhomogeneous strain fields originating at the intersection of various kinds of nanoscopic monoclinic domains that form accordion-like structures, as shown in Figure 9(c). Following this work, Cheema et al. [61] have reported well textured ultrathin layers of HZO directly grown on Si using atomic layer deposition. The authors report that the films below 2.5 nm are well-oriented, perhaps predominantly along (100) (and definitely not (111)- from pole figures). These films exhibit enhanced acentric rhombic distortions of oxygen tetrahedra at a local scale [18], resulting in large orthorhombic d$_{(111)}$. These enhanced lattice distortions at smaller thicknesses are similar in nature to the r-phase films grown using PLD [30]. Also, in this work, it seems likely that surface energy and inhomogenous strain fields are responsible for the stabilization of an o-phase. Apart from Si (100), there are not many substrates that offer compressive strain for a cube-on-cube growth of HZO (001) layers. So, the effect of compressive strain on any phase-stabilization of the (001) layers is currently elusive, and thus makes for a future prospect.

## 4. Conclusions and outlook

Pure polar r-phase is stabilized in HZO layers using a combination of compressive strain with (111) orientation of the films. A direct way of engineering the r-phase is to grow HZO on (0001)-oriented hexagonal substrates such as GaN, sapphire, or cubic Si (111) surfaces. These substrates provide the template necessary to force HZO to grow along (111), while imposing an in-plane compressive strain. Since the $P_r$ for HZO on these substrates is low ($<2\,\mu C/cm^2$), the depolarization effects are also not important for destabilizing this phase.

Another way of engineering the r-phase is to utilize surface-energy mediated growth modes, which orient the film along (111) given the low-surface energy of these faces. Compressive strain due to thermal expansion coefficient mismatch, in addition to the hydrostatic pressure provided by the nano-domain structure, stabilizes the single r-phase. This happens for HZO grown on perovskites such as STO and DSO with LSMO under tensile strain. On these substrates, HZO layers exhibit large $P_r$ (up to 34 $\mu C/cm^2$), and thus stabilization of such a phase would require efficient screening mechanisms. Tensile-straining LSMO creates an oxygen-deficient layer at the LSMO-HZO interface that stabilizes the (111) orientation, the r-phase, and the large $P_r$ associated with it. Polar o-phase in HZO seems to have been recently observed in ultrathin layers of HZO grown on Si (100) using atomic layer deposition, although whether it is unique and stand-alone is not clear [61]. Stabilizing and studying this phase preferentially through appropriate selection of substrate and strain-engineering remains a prospective study.

## Acknowledgments

PN would like to acknowledge funding from European Union's Horizon 2020 research and innovation program under Marie Sklodowska-Curie grant agreement #794954 (nickname: FERHAZ). Y.W. acknowledges a China Scholarship Council grant and a Van Gogh travel grant.